\documentclass{emulateapj}
\usepackage{psfig,natbib,amsmath}

\def\sn{SN~2008es}
\def\kms{km~s$^{-1}$}
\def\swift{\emph{Swift}}
\def\arcmin{\hbox{$^\prime$}}
\def\arcsec{\hbox{$^{\prime\prime}$}}

\def\jhk{{\em J}, {\em H}, and {\em K$_s$}}
\newcommand{\bvri}{\protect\hbox{$BV\!RI$}}
\newcommand{\citepeg}[1]{\citep[{e.g.,}][]{#1}}

\shorttitle{Luminous Type II-L \sn}

\shortauthors{Miller et al.}

\begin{document}

\title{The Exceptionally Luminous Type II-Linear Supernova 2008es}

\def\berk{1}
\def\glast{2}
\def\sloan{3}

\author{A. A. Miller\altaffilmark{\berk}, 
R. Chornock\altaffilmark{\berk},   
D. A. Perley\altaffilmark{\berk},  
M. Ganeshalingam\altaffilmark{\berk}, 
W. Li\altaffilmark{\berk},
N. R. Butler\altaffilmark{\berk,\glast}, 
J. S. Bloom\altaffilmark{\berk,\sloan}, 
N. Smith\altaffilmark{\berk},
M. Modjaz\altaffilmark{\berk}, 
D. Poznanski\altaffilmark{\berk}, 
A. V. Filippenko\altaffilmark{\berk}, 
C. V. Griffith\altaffilmark{\berk}, 
J. H. Shiode\altaffilmark{\berk}, 
and 
J. M. Silverman\altaffilmark{\berk}}

\altaffiltext{\berk}{Department of Astronomy, University of
California, Berkeley, CA 94720-3411. Email: amiller@astro.berkeley.edu}
\altaffiltext{\glast}{GLAST Fellow.}

\altaffiltext{\sloan}{Sloan Research Fellow.}

\begin{abstract}

We report on our early photometric and spectroscopic observations of
the extremely luminous Type II supernova (SN) 2008es. With an observed
peak optical magnitude of $m_V = 17.8$ and at a redshift $z = 0.213$,
\sn\ had a peak absolute magnitude of $M_V$ = $-22.3$, making it the
second most luminous SN ever observed. The photometric evolution of
\sn\ exhibits a fast decline rate ($\sim$0.042 mag d$^{-1}$),
similar to the extremely luminous Type II-L SN~2005ap.  We show that
\sn\ spectroscopically resembles the luminous Type II-L SN~1979C.
Although the spectra of \sn\ lack the narrow and intermediate-width
line emission typically associated with the interaction of a SN with
the circumstellar medium of its progenitor star, we argue that the
extreme luminosity of \sn\ is powered via strong interaction with a
dense, optically thick circumstellar medium. The integrated bolometric
luminosity of \sn\ yields a total radiated energy at ultraviolet and
optical wavelengths of $\ga$10$^{51}$ ergs.  Finally, we examine the
apparently anomalous rate at which the Texas Supernova Search has
discovered rare kinds of supernovae, including the five most luminous
supernovae observed to date, and find that their results are
consistent with those of other modern SN searches.

\end{abstract}


\section{Introduction}

Wide-field synoptic optical imaging surveys are continuing to probe
the parameter space of time-variable phenomena with increasing depth
and temporal coverage
\citepeg{2004ApJ...611..418B,2006MNRAS.371.1681M,2008MNRAS.386..887B},
unveiling a variety of transients ranging from the common
\citep{2008ApJ...682.1205R} to the unexplained (e.g.,
\citealt{bdt+08}). Untargeted (``blind'') synoptic wide-field imaging
surveys, such as the Texas Supernova Search (TSS; \citealt{quimbyTSS})
conducted with the ROTSE-III 0.45-m telescope \citep{akerlof03}, have
uncovered a large number of rare transients, including the four most
luminous supernovae (SNe) observed to date: SN~2005ap
\citep{quimby05ap}, SN~2008am \citep{ATEL.1389}, SN~2006gy
\citep{ofek06gy, smith07-2006gy, smith08-2006gy}, and SN~2006tf
\citep{smith06tf}.  Observations of these very luminous events are
starting to allow the detailed physical study of the extrema in
core-collapse SNe. They appear to be powered in part by their
interaction with a highly dense circumstellar medium (CSM; see
\citealt{smith06tf}, and references therein), though other
possibilities have been advanced. Clearly, the discovery of more such
events would allow an exploration of the variety of the phenomenology
as related to the diversity of progenitors and CSM.

Recently, the TSS discovered yet another luminous transient on 2008
Apr.  26.23 (UT dates are used throughout this paper), which they
suggested was a variable active galactic nucleus at a redshift $z =
1.02$ \citep{ATEL.1515}. \citet{ATEL.1524} then hypothesized that the
transient was a flare from the tidal disruption of a star by a
supermassive black hole. \citet{ATEL.1576} first identified \sn\ as
potentially an extremely luminous Type II SN (see also
\citealt{2008ATel.1578....1G}), and we later definitively confirmed
this with further spectroscopic observations \citep{ATEL.1644}; the
event was assigned the name \sn\ by the IAU
\citep{2008CBET.1462....1C}. It is located at $\alpha$ =
11$^h$56$^m$49.06$^s$, $\delta$ = +54$^o$27$\arcmin$24.77$\arcsec$
(J2000.0).

Here we present our analysis of SN 2008es, which is classified as a
Type II-Linear (II-L) SN based on the observed linear (in mag) decline
in the photometric light curve \citep{barbon79,doggett85}. At $z$ =
0.213, \sn\ has a peak optical magnitude of $M_V = -22.3$, among SNe
second only to SN~2005ap. Aside from the extreme luminosity, \sn\ is
of great interest since detailed ultraviolet (UV) through infrared
(IR) observations provide a unique opportunity to study the mass-loss
properties of an evolved post-main sequence massive star via its
interaction with the surrounding dense CSM. A similar analysis of \sn\
has been presented by \citet{gezari08}.

The outline of this paper is as follows. We present our observations
in $\S$2, and the photometric and spectroscopic analyses of this and
public (NASA) data in $\S$3 and $\S$4, respectively. A discussion is
given in $\S$5, and our conclusions are summarized in
$\S$6. Throughout this paper we adopt a concordance cosmology of $H_0$
= 70 \kms\ Mpc$^{-1}$, $\Omega_{\rm M} = 0.3$, and $\Omega_\Lambda =
0.7$.

\section{Observations}

\begin{figure}[h]
\centerline{\psfig{file=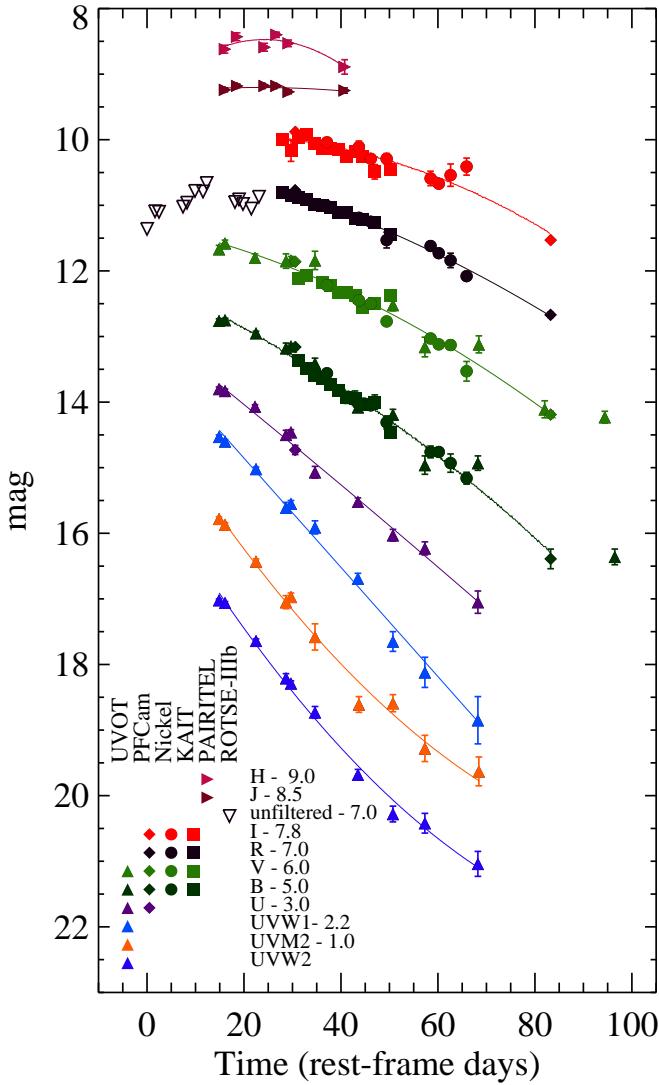,width=3.5in,angle=0}}
\caption{Observed UV-optical-NIR light curves of \sn. The data have
not been corrected for Galactic or host-galaxy extinction. We include
ROTSE-IIIb unfiltered observations from the literature (open symbols;
\citealt{gezari08}), as well as our optical-NIR observations (filled
symbols; KAIT, Nickel, PFCam, and PAIRITEL) and space-based UVOT
observations from \swift\ (filled triangles). We adopt the discovery
date of \sn, 2008 Apr. 26.23, as ``day 0'' for this SN. Low-order
polynomial fits to each band have been overplotted to help guide the
eye.}
\label{ourphot}
\end{figure}

Here we present our ground-based optical and near-infrared (NIR)
photometry and optical spectroscopy, along with space-based \swift\
UV, optical, and X-ray observations. NIR photometry of \sn\ was
obtained simultaneously in \jhk\ with the Peters Automated Infrared
Imaging Telescope (PAIRITEL; \citealt{bloom06}) beginning 2008 May 16.
To improve the photometric signal-to-noise ratio (S/N), we stacked
images over multiple nights. For the {\em K$_s$} images the S/N of the
SN remained low, despite the stacks made over multiple epochs, and
therefore we do not include these data in our subsequent
analysis. Aperture photometry, using a custom pipeline, was used to
measure the {\em J} and {\em H} photometry of the isolated SN
calibrated to the 2MASS catalog (see \citealt{bloom080319b}). The
resulting light curves are presented in Figure~\ref{ourphot}. The
final PAIRITEL photometry is reported in Table~\ref{tbl-pairitel}.

\begin{deluxetable}{ccccr}
\tablecaption{PAIRITEL Observations of \sn\label{tbl-pairitel}}
\tablecolumns{5}
\tablewidth{3.0in}
\tablehead{\colhead{$t_{\rm mid}$\tablenotemark{a}} & \colhead{Obs.
window\tablenotemark{b}} & \colhead{Filter} & \colhead{Exp. time} &
\colhead{Mag\tablenotemark{c}} \\ 
\colhead{(day)} & \colhead{(day)} & & \colhead{(sec)} & }
\startdata
4602.24 & 2.06 & J & 2895.91 & 17.74$\pm$0.03 \\
4605.28 & 6.09 & J & 3869.06 & 17.68$\pm$0.03 \\
4612.17 & 2.09 & J & 4606.78 & 17.68$\pm$0.02 \\
4615.15 & 6.08 & J & 6458.90 & 17.68$\pm$0.02 \\
4618.19 & 2.19 & J & 6129.29 & 17.77$\pm$0.03 \\
4632.17 & 6.04 & J & 4944.24 & 17.75$\pm$0.04 \\
4602.24 & 2.06 & H & 2880.22 & 17.62$\pm$0.06 \\
4605.27 & 6.09 & H & 3751.34 & 17.43$\pm$0.06 \\
4612.17 & 2.09 & H & 4598.93 & 17.59$\pm$0.06 \\
4615.15 & 6.08 & H & 6451.06 & 17.40$\pm$0.04 \\
4618.19 & 2.19 & H & 6074.35 & 17.53$\pm$0.05 \\
4632.17 & 6.04 & H & 4865.76 & 17.89$\pm$0.11 \\
\enddata
\tablecomments{PAIRITEL observations were stacked over multiple epochs to increase
the S/N.}
\tablenotetext{a}{Mid-point between the first and last exposures in a single stacked
image, reported as JD$-$2,450,000.}
\tablenotetext{b}{Time between the first and last exposures in a single stacked image.}
\tablenotetext{c}{Observed value; not corrected for Galactic extinction.}
\end{deluxetable}

Optical photometry of \sn\ was obtained in \bvri\ with the Katzman
Automatic Imaging Telescope (KAIT; \citealt{filippenkoLOSS}) and the
1-m Nickel telescope at Lick Observatory beginning 2008 May
30. Point-spread function (PSF)-fitting photometry was performed on
the SN and several comparison stars using the {\tt IRAF/DAOPHOT}
package \citep{1987PASP...99..191S:stetson} and transformed into the
Johnson-Cousins system. Calibrations for the field were obtained with
the Nickel telescope on three photometric nights. The final photometry
from KAIT and the Nickel telescope is given in Tables~\ref{tbl-kait}
and~\ref{tbl-nickel}, respectively.

\begin{deluxetable}{cccr}
\tablecaption{KAIT Observations of \sn\label{tbl-kait}}
\tablecolumns{4}
\tablewidth{2.2in}
\tablehead{\colhead{$t_{\rm obs}$\tablenotemark{a}} & \colhead{Filter} &
\colhead{Exp. time} & \colhead{Mag\tablenotemark{b}} \\ 
\colhead{(day)} & & \colhead{(sec)} & }
\startdata
      4620.72 & B & 360.00 & 18.36$\pm$0.04 \\
      4622.73 & B & 360.00 & 18.49$\pm$0.05 \\
      4624.72 & B & 360.00 & 18.60$\pm$0.04 \\
      4626.72 & B & 360.00 & 18.64$\pm$0.06 \\
      4628.72 & B & 360.00 & 18.73$\pm$0.07 \\
      4630.72 & B & 360.00 & 18.83$\pm$0.06 \\
      4632.74 & B & 360.00 & 18.93$\pm$0.09 \\
      4634.77 & B & 360.00 & 18.94$\pm$0.11 \\
      4636.73 & B & 360.00 & 19.03$\pm$0.09 \\
      4639.70 & B & 360.00 & 19.00$\pm$0.11 \\
      4643.74 & B & 360.00 & 19.46$\pm$0.09 \\
      4620.72 & V & 300.00 & 18.12$\pm$0.06 \\
      4622.73 & V & 300.00 & 18.07$\pm$0.03 \\
      4626.72 & V & 300.00 & 18.17$\pm$0.04 \\
      4628.72 & V & 300.00 & 18.22$\pm$0.04 \\
      4630.72 & V & 300.00 & 18.33$\pm$0.04 \\
      4632.74 & V & 300.00 & 18.33$\pm$0.06 \\
      4634.77 & V & 300.00 & 18.38$\pm$0.09 \\
      4636.73 & V & 300.00 & 18.55$\pm$0.03 \\
      4639.70 & V & 300.00 & 18.50$\pm$0.04 \\
      4643.74 & V & 300.00 & 18.37$\pm$0.05 \\
      4616.74 & R & 300.00 & 17.81$\pm$0.03 \\
      4618.76 & R & 300.00 & 17.85$\pm$0.04 \\
      4620.72 & R & 300.00 & 17.88$\pm$0.04 \\
      4622.73 & R & 300.00 & 17.91$\pm$0.03 \\
      4624.72 & R & 300.00 & 17.99$\pm$0.04 \\
      4626.72 & R & 300.00 & 18.00$\pm$0.03 \\
      4628.72 & R & 300.00 & 18.03$\pm$0.04 \\
      4630.72 & R & 300.00 & 18.11$\pm$0.05 \\
      4632.74 & R & 300.00 & 18.11$\pm$0.03 \\
      4634.77 & R & 300.00 & 18.20$\pm$0.04 \\
      4636.73 & R & 300.00 & 18.21$\pm$0.05 \\
      4639.70 & R & 300.00 & 18.26$\pm$0.05 \\
      4643.74 & R & 300.00 & 18.44$\pm$0.06 \\
      4616.74 & I & 300.00 & 17.74$\pm$0.07 \\
      4618.76 & I & 300.00 & 17.92$\pm$0.16 \\
      4620.72 & I & 300.00 & 17.72$\pm$0.05 \\
      4622.73 & I & 300.00 & 17.67$\pm$0.05 \\
      4624.72 & I & 300.00 & 17.81$\pm$0.05 \\
      4626.72 & I & 300.00 & 17.88$\pm$0.04 \\
      4628.72 & I & 300.00 & 17.88$\pm$0.05 \\
      4630.72 & I & 300.00 & 17.90$\pm$0.07 \\
      4632.74 & I & 300.00 & 18.01$\pm$0.06 \\
      4634.77 & I & 300.00 & 17.93$\pm$0.07 \\
      4636.73 & I & 300.00 & 18.01$\pm$0.08 \\
      4639.70 & I & 300.00 & 18.24$\pm$0.11 \\
      4643.74 & I & 300.00 & 18.20$\pm$0.09 \\
\enddata
\tablenotetext{a}{Exposure mid-point, reported as JD$-$2,450,000.}
\tablenotetext{b}{Observed value; not corrected for Galactic extinction.}
\end{deluxetable}

\begin{deluxetable}{cccr}
\tablecaption{Nickel Observations of \sn\label{tbl-nickel}}
\tablecolumns{4}
\tablewidth{2.2in}
\tablehead{\colhead{$t_{\rm obs}$\tablenotemark{a}} & \colhead{Filter} &
\colhead{Exp. time} & \colhead{Mag\tablenotemark{b}} \\ 
\colhead{(day)} & & \colhead{(sec)} & }
\startdata
      4627.75 & B & 360.00 & 18.56$\pm$0.02 \\
      4635.75 & B & 360.00 & 19.04$\pm$0.07 \\
      4638.71 & B & 360.00 & 19.05$\pm$0.03 \\
      4642.67 & B & 360.00 & 19.31$\pm$0.10 \\
      4653.67 & B & 360.00 & 19.76$\pm$0.09 \\
      4655.71 & B & 360.00 & 19.76$\pm$0.04 \\
      4658.67 & B & 360.00 & 19.93$\pm$0.14 \\
      4662.67 & B & 360.00 & 20.16$\pm$0.09 \\
      4627.75 & V & 300.00 & 18.22$\pm$0.03 \\
      4635.75 & V & 300.00 & 18.45$\pm$0.04 \\
      4638.71 & V & 300.00 & 18.50$\pm$0.03 \\
      4642.67 & V & 300.00 & 18.77$\pm$0.04 \\
      4653.67 & V & 300.00 & 19.03$\pm$0.04 \\
      4655.71 & V & 300.00 & 19.12$\pm$0.04 \\
      4658.67 & V & 300.00 & 19.13$\pm$0.07 \\
      4662.67 & V & 300.00 & 19.53$\pm$0.15 \\
      4627.75 & R & 300.00 & 18.01$\pm$0.02 \\
      4635.75 & R & 300.00 & 18.19$\pm$0.04 \\
      4638.71 & R & 300.00 & 18.25$\pm$0.03 \\
      4642.67 & R & 300.00 & 18.53$\pm$0.12 \\
      4653.67 & R & 300.00 & 18.62$\pm$0.04 \\
      4655.71 & R & 300.00 & 18.73$\pm$0.03 \\
      4658.67 & R & 300.00 & 18.84$\pm$0.11 \\
      4662.67 & R & 300.00 & 19.08$\pm$0.05 \\
      4627.75 & I & 300.00 & 17.79$\pm$0.04 \\
      4635.75 & I & 300.00 & 17.86$\pm$0.09 \\
      4638.71 & I & 300.00 & 18.04$\pm$0.04 \\
      4642.67 & I & 300.00 & 18.04$\pm$0.08 \\
      4653.67 & I & 300.00 & 18.34$\pm$0.11 \\
      4655.71 & I & 300.00 & 18.42$\pm$0.07 \\
      4658.67 & I & 300.00 & 18.29$\pm$0.17 \\
      4662.67 & I & 300.00 & 18.16$\pm$0.13 \\
\enddata
\tablenotetext{a}{Exposure mid-point, reported as JD$-$2,450,000.}
\tablenotetext{b}{Observed value; not corrected for Galactic extinction.}
\end{deluxetable}

Additional optical photometry was obtained in $U$\bvri\ on 2008 June
02 and \bvri\ on 2008 Aug. 05 with PFCam on the 3-m Shane telescope at
Lick Observatory. The data were reduced using standard techniques and
aperture photometry was used to extract the SN flux. \bvri\
calibrations were done using the same stars as those used for the
Nickel and KAIT observations. Calibrations for the $U$ band were not
obtained with the Nickel telescope. Therefore, we convert the 
Sloan Digital Sky Survey (SDSS) colors of stars in the field to the 
$U$ band with the color
transformations of \citet{jester2005}, and use these stars for our
$U$-band calibration. The final PFCam photometry is reported in
Table~\ref{tbl-pfcam}.

\begin{deluxetable}{cccr}
\tablecaption{PFCam Observations of \sn\label{tbl-pfcam}}
\tablecolumns{4}
\tablewidth{2.2in}
\tablehead{\colhead{$t_{\rm obs}$\tablenotemark{a}} & \colhead{Filter} &
\colhead{Exp. time} & \colhead{Mag\tablenotemark{b}} \\ 
\colhead{(day)} & & \colhead{(sec)} & }
\startdata
      4619.81 & U & 1230.00 & 17.73$\pm$0.07 \\
      4619.83 & B & 630.00 & 18.16$\pm$0.01 \\
      4683.70 & B & 360.00 & 21.39$\pm$0.15 \\
      4619.83 & V & 930.00 & 17.86$\pm$0.01 \\
      4683.71 & V & 360.00 & 20.19$\pm$0.05 \\
      4619.79 & R & 930.00 & 17.77$\pm$0.01 \\
      4683.68 & R & 600.00 & 19.67$\pm$0.04 \\
      4619.85 & I & 1230.00 & 17.63$\pm$0.01 \\
      4683.70 & I & 360.00 & 19.28$\pm$0.03 \\
\enddata
\tablenotetext{a}{Exposure mid-point, reported as JD-2450000.}
\tablenotetext{b}{Observed value; not corrected for Galactic extinction.}
\end{deluxetable}

The \swift\ satellite observed \sn\ during 13 epochs between 2008 May
14 and Aug. 21.  We downloaded the Ultraviolet/Optical Telescope
(UVOT; \citealt{roming05}) data from the \swift\ data archive and
analyzed the Level 2 sky image data in $U$, $B$, and $V$ according to
the photometry calibration and recipe by \citet{li06}.  The \swift\ UV
filters ($UVW1$, $UVM1$, and $UVW2$) were reduced following
\citet{poole2008}.  The final \swift\ UVOT photometry is reported in
Table~\ref{tbl-uvot}.

\begin{deluxetable}{cccr}
\tabletypesize{\small}
\tablecaption{UVOT Observations of \sn}
\tablecolumns{4}
\tablewidth{2.2in}
\tablehead{\colhead{$t_{\rm obs}$\tablenotemark{a}} & \colhead{Filter} &
\colhead{Exp. time} & \colhead{Mag\tablenotemark{b}} \\ 
\colhead{(day)} & \colhead{} & \colhead{(sec)} & \colhead{}}
\startdata
      4600.75 & UVW2 & 1585.70 & 17.02$\pm$0.02 \\
      4602.25 & UVW2 & 1802.20 & 17.06$\pm$0.02 \\
      4610.00 & UVW2 & 2119.30 & 17.64$\pm$0.03 \\
      4617.50 & UVW2 & 542.40 & 18.21$\pm$0.07 \\
      4618.75 & UVW2 & 1731.10 & 18.29$\pm$0.04 \\
      4624.75 & UVW2 & 487.70 & 18.73$\pm$0.09 \\
      4635.50 & UVW2 & 1664.50 & 19.68$\pm$0.08 \\
      4644.25 & UVW2 & 1593.40 & 20.28$\pm$0.12 \\
      4652.25 & UVW2 & 1317.80 & 20.42$\pm$0.15 \\
      4665.50 & UVW2 & 1765.70 & 21.04$\pm$0.19 \\
      4600.75 & UVM2 & 1043.20 & 16.78$\pm$0.03 \\
      4602.25 & UVM2 & 1148.40 & 16.87$\pm$0.03 \\
      4610.00 & UVM2 & 1523.60 & 17.43$\pm$0.04 \\
      4617.50 & UVM2 & 338.60 & 18.05$\pm$0.10 \\
      4618.75 & UVM2 & 1110.60 & 17.97$\pm$0.06 \\
      4624.75 & UVM2 & 150.30 & 18.58$\pm$0.20 \\
      4635.75 & UVM2 & 1149.50 & 19.61$\pm$0.12 \\
      4644.25 & UVM2 & 1081.80 & 19.59$\pm$0.13 \\
      4652.25 & UVM2 & 938.10 & 20.28$\pm$0.20 \\
      4665.75 & UVM2 & 1071.50 & 20.63$\pm$0.22 \\
      4600.75 & UVW1 & 792.40 & 16.78$\pm$0.03 \\
      4602.25 & UVW1 & 900.70 & 16.85$\pm$0.03 \\
      4610.00 & UVW1 & 1058.30 & 17.27$\pm$0.03 \\
      4617.50 & UVW1 & 270.70 & 17.86$\pm$0.08 \\
      4618.75 & UVW1 & 865.10 & 17.80$\pm$0.05 \\
      4624.75 & UVW1 & 243.70 & 18.16$\pm$0.10 \\
      4635.50 & UVW1 & 831.80 & 18.94$\pm$0.08 \\
      4644.25 & UVW1 & 796.10 & 19.90$\pm$0.15 \\
      4652.25 & UVW1 & 790.00 & 20.37$\pm$0.23 \\
      4665.50 & UVW1 & 1048.30 & 21.10$\pm$0.36 \\
      4600.75 & U & 395.80 & 16.80$\pm$0.03 \\
      4602.25 & U & 449.90 & 16.83$\pm$0.03 \\
      4609.75 & U & 399.60 & 17.07$\pm$0.03 \\
      4617.50 & U & 134.90 & 17.50$\pm$0.07 \\
      4618.75 & U & 431.90 & 17.46$\pm$0.04 \\
      4624.75 & U & 121.50 & 18.07$\pm$0.09 \\
      4635.50 & U & 415.40 & 18.52$\pm$0.06 \\
      4644.25 & U & 397.40 & 19.03$\pm$0.09 \\
      4652.25 & U & 386.70 & 19.23$\pm$0.10 \\
      4665.50 & U & 523.60 & 20.05$\pm$0.17 \\
      4600.75 & B & 395.80 & 17.76$\pm$0.03 \\
      4602.25 & B & 449.90 & 17.75$\pm$0.03 \\
      4610.00 & B & 527.80 & 17.95$\pm$0.03 \\
      4617.50 & B & 135.00 & 18.18$\pm$0.08 \\
      4618.75 & B & 431.90 & 18.15$\pm$0.04 \\
      4624.75 & B & 121.60 & 18.41$\pm$0.08 \\
      4635.50 & B & 415.40 & 19.08$\pm$0.06 \\
      4644.25 & B & 397.50 & 19.19$\pm$0.08 \\
      4652.25 & B & 328.60 & 19.96$\pm$0.14 \\
      4665.50 & B & 523.70 & 19.93$\pm$0.11 \\
      4699.75 & B & 4150.20 & 21.36$\pm$0.12 \\
      4600.75 & V & 395.70 & 17.67$\pm$0.06 \\
      4602.25 & V & 449.90 & 17.58$\pm$0.05 \\
      4609.75 & V & 399.50 & 17.80$\pm$0.06 \\
      4617.50 & V & 134.90 & 17.85$\pm$0.11 \\
      4618.75 & V & 432.00 & 17.84$\pm$0.06 \\
      4624.75 & V & 81.00 & 17.84$\pm$0.14 \\
      4635.75 & V & 415.40 & 18.41$\pm$0.08 \\
      4644.25 & V & 397.60 & 18.52$\pm$0.09 \\
      4652.25 & V & 328.50 & 19.16$\pm$0.15 \\
      4665.75 & V & 398.90 & 19.12$\pm$0.13 \\
      4682.22 & V & 1876.40 & 20.11$\pm$0.13 \\
      4697.25 & V & 4316.10 & 20.23$\pm$0.09 \\
\enddata
\tablenotetext{a}{Exposure mid-point, reported as JD$-$2,450,000.}
\tablenotetext{b}{Observed value; not corrected for Galactic extinction.}
\label{tbl-uvot}
\end{deluxetable}

Simultaneous observations of \sn\ occurred with the \swift\ X-ray
Telescope (XRT; \citealt{burrows05}), for which we confirm a
nondetection of X-ray emission (see also \citealt{ATEL.1524}). To
place a limit on the source flux we assume a power-law spectrum with a
photon index $\Gamma = 2$, absorbed by elements associated with
Galactic \ion{H}{1}. We took an extraction region of radius 64 pixels
($\sim$2.5$\arcmin$) and fit the PSF model around the centroid of the
optical emission. We obtain a 3$\sigma$ limiting flux of $9 \times
10^{-15}$ erg~cm$^{-2}$~s$^{-1}$ in the 0.3--10 keV band for a total
exposure of 54.7 ks.  This represents an upper limit to the X-ray
luminosity of \sn\ of $\sim 1.2 \times 10^{42}$ ergs~s$^{-1}$.

\begin{figure}
\centerline{\psfig{file=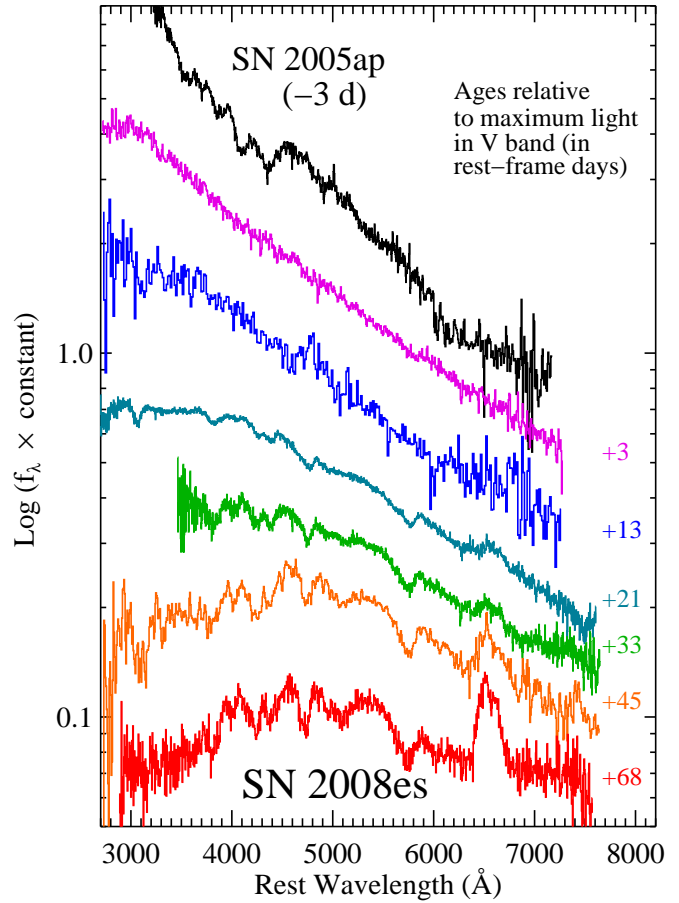,width=3.5in,angle=0}}
\caption{Spectral evolution of \sn. Spectra of \sn\ (in color) are
labeled with their ages in the rest frame of the supernova ($z =
0.213$) relative to the observed $V$-band maximum on May 13.3.  The
spectra become progressively redder as the SN ages and broad P-Cygni
spectral features become more prominent with time.  By +68~d, a broad
emission feature of H$\alpha$ is clearly present.  The top spectrum
(in black) is the earliest spectrum of SN~2005ap, the most luminous
observed SN \citep{quimby05ap}.}
\label{specevo}
\end{figure}

We obtained spectra of \sn\ on 2008 May 16.3, 2008 May 29.3, and 2008
July 7.3 using the Kast spectrograph on the Lick 3-m telescope
\citep{kastref}. Additional spectra were obtained on 2008 June 7.4 and
2008 Aug. 3.3 using the Low Resolution Imaging Spectrometer on the
Keck I 10-m telescope \citep{oke95} and on 2008 June 21.2 and 2008
June 23.2 using the R.\ C.\ Spectrograph on the Kitt Peak 4-m
telescope, following the approval of our request for Kitt Peak
Director's Discretionary Time. The spectra were extracted and
calibrated following standard procedures (e.g., \citealt{matheson00}).
Clouds were present during several of the observations, making the
absolute flux scales unreliable.  The spectrograph slit was placed at
the parallactic angle, so the relative spectral shapes should be
accurate \citep{fil82}, with the exception of the Kitt Peak spectra,
which have a small amount of second-order light contamination at
wavelengths longward of $\sim$8000~\AA.  The full \sn\ spectral
sequence is plotted in Figure~\ref{specevo}.  The two Kitt Peak
spectra show little evolution in the two days that separate them and
have been combined to increase the S/N (day 33 in
Figure~\ref{specevo}).  A log of our spectroscopic observations is
presented in Table~\ref{specobstab}.

\begin{deluxetable*}{lccccccc}
\tablecaption{Log of Spectroscopic Observations}
\tablewidth{0pt}
\tablehead{\colhead{Age\tablenotemark{a}} & \colhead{UT Date} &
\colhead{Instrument\tablenotemark{b}} &
\colhead{Range} & \colhead{Exp. Time} & \colhead{Seeing} &
\colhead{Airmass} & \colhead{Photometric?} \\
\colhead{(days)} &  &  & \colhead{(\AA)}
  & \colhead{(s)} & \colhead{(\arcsec)} &  & \colhead{(y/n)} }
\startdata
3 & 2008-05-16.345 & Kast & 3300--8830 &  1500 & 2.4 & 1.3 & y \\
13 & 2008-05-29.253 & Kast & 3300--8820 & 1800 & 2.3 & 1.1 & n \\
21 & 2008-06-07.399 & LRIS & 3200--9230 & 1200 & 1.6 & 2.0 & y \\
33 & 2008-06-21.194 & RC & 4200--9280 & 1800 & 2.9 & 1.3 & y \\
34 & 2008-06-23.189 & RC & 4200--9280 & 2400 & 1.8 & 1.3 & y \\
45 & 2008-07-07.252 & Kast & 3300--9280 & 4200 & 2.5 & 1.6 & y \\
68 & 2008-08-03.254 & LRIS & 3500--9190 & 877 & 0.9 & 2.0 & n \\
\enddata
\tablenotetext{a}{Age in rest-frame days relative to the observed
  $V$-band maximum on 2008 May 13.3.}
\tablenotetext{b}{Kast = Kast spectrograph on Lick 3-m telescope. LRIS
  = Low Resolution Imaging Spectrometer on Keck-I 10-m telescope.  RC
  = R.\ C.\ Spectrograph on Kitt Peak 4-m telescope.}
\label{specobstab}
\end{deluxetable*}

We searched our spectra for the possible presence of narrow lines and
were unable to positively identify any, either in emission or
absorption. Therefore, without a detection of the host galaxy, we
determine the redshift of \sn\ directly from the SN spectrum. As the
SN aged, broad P-Cygni spectral features appeared, including an
emission feature near 7900~\AA\ that we identify as H$\alpha$ near a
redshift of 0.2.  Other spectral features are consistent with a Type
II SN at about that redshift.

\begin{figure}
\centerline{\psfig{file=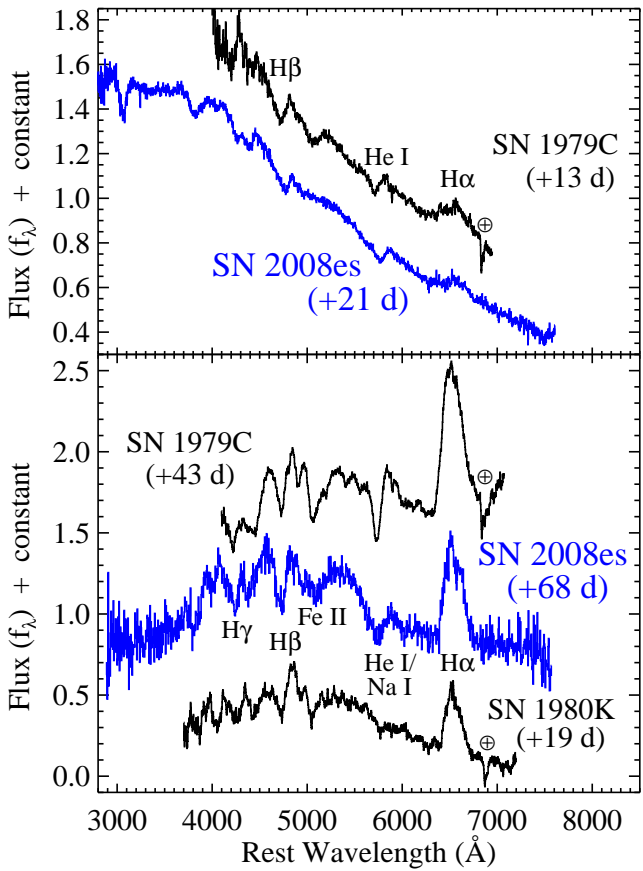,width=3.5in,angle=0}}
\caption{Spectral comparisons of \sn\ to two SNe II-L.  All spectra
are labeled with their respective ages in rest-frame days relative to
maximum light, and prominent spectral features are identified.  The
zero point of the flux scale is accurate for the \sn\ spectra (in
blue), while the comparison spectra (in black) are vertically offset.
The top panel shows a comparison at an early epoch of \sn\ to SN~1979C
from 1979 Apr. 28 \citep{branch81}.  The bottom panel shows a
comparison at a later epoch to SN~1979C from 1979 May 28
\citep{branch81} and SN~1980K (spectra from 1980 Nov. 15 and 17
combined; \citealt{uomoto86}).  Telluric absorption bands in the SNe
1979C and 1980K spectra are marked with a $\earth$ symbol.}
\label{speccomp}
\end{figure}

To get a more accurate redshift, we identified two similar reference
spectra of the Type II-L SNe 1979C and 1980K from the literature
\citep{branch81,uomoto86}.  These spectral comparisons are shown in
the bottom panel of Figure~\ref{speccomp}.  We used the SuperNova
IDentification code of \citet{snid07} to cross-correlate the day 68
spectrum of \sn\ with the two reference spectra and derived a
weighted-average $z = 0.213 \pm 0.006$, which we adopt throughout this
paper.  This redshift agrees with a fit to the broad H$\alpha$
emission line in the day 68 spectrum, which yields a flux centroid of
$z = 0.210$.  The \sn\ redshift determined using this method is
correlated with the expansion velocity, but we assume that the bias
due to this effect is small given that all three SNe have H$\alpha$
emission lines of a similar width at late times.

\section{Photometric Results}

Our dataset provides excellent broad-band coverage of \sn\ from the UV
to the NIR, which allows us to model changes in the spectral energy
distribution (SED). In order to sample each photometric band onto a
single set of common epochs, we fit low-order polynomials to each
light curve, which we then interpolate onto a common grid. NIR
observations were only included on or around epochs where we detected
the SN.

\begin{figure}
\centerline{\psfig{file=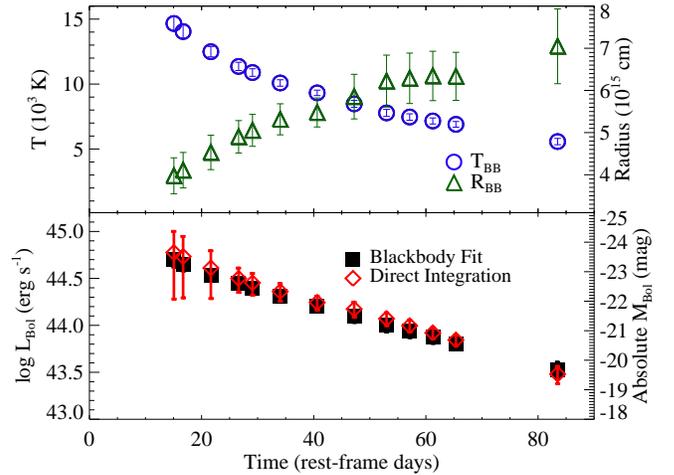,width=3.5in,angle=0}}
\caption{Photospheric and bolometric luminosity evolution of \sn. Top:
temperature evolution of \sn\ based on BB fits (open circles) and
inferred radius (open triangles).  The error bars on the radius are
likely to be slightly overestimated because the errors on the BB
temperature and luminosity are correlated.  Bottom: bolometric
luminosity of \sn\ derived via two independent methods, BB modeling
(closed squares) and integration of the total UV+optical (+NIR where
available) flux (open diamonds). The two methods agree to within $\la$
20\%.}
\label{bol_lum}
\end{figure}

We create an SED at each of the common epochs and fit a
single-component blackbody (BB) to the data following the procedure
described by \citet{modjaz08d}. Prior to the SED/BB fits we add a
systematic term to the uncertainty in the photometric measurement in
each band. This term is added in quadrature to the statistical
uncertainty, and for the NIR ({\em JH}) is equal to 2\%, in the
optical (\bvri) we adopt 3\%, while for $U$ band we adopt 10\%, and
the adopted UV ($UVW1$, $UVM2$, $UVW2$) systematic uncertainty is
20\%. Across all epochs of the BB fits the total $\chi^2$ per degree
of freedom is $45.6/65$, which suggests that our systematic terms may
be slightly overestimated.  We assume no host-galaxy extinction (for
further details see $\S$5), and correct our measurements for the
modest amount of Galactic reddening $E(B-V)$ = 0.011 mag
\citep{schlegel98}. From the SED/BB fits we derive the temperature,
radius, and luminosity of the SN as a function of time, as shown in
Figure~\ref{bol_lum}.  The temperature and luminosity decrease while
the radius increases with time, as expected for an expanding and
cooling SN photosphere.

In order to determine the time of $V$-band maximum light, which at the
redshift of this SN roughly corresponds to rest-frame $B$, we convert
the early time $g$, $r$, and $i'$ photometry taken with the Palomar
1.5-m telescope from \citet{gezari08} to $VRI$ using the color
equations from \citet{jester2005}. Following a quadratic fit to these
data, we find that the observed $V$-band maximum occurred on
$\sim$2008 May 13, $\sim$15 rest-frame days after discovery. The
observed peak is $V$ = 17.8 mag, which corresponds to $M_V$ = $-22.3$
mag, with a scatter about our fit of 0.04 mag.

\begin{figure}
\centerline{\psfig{file=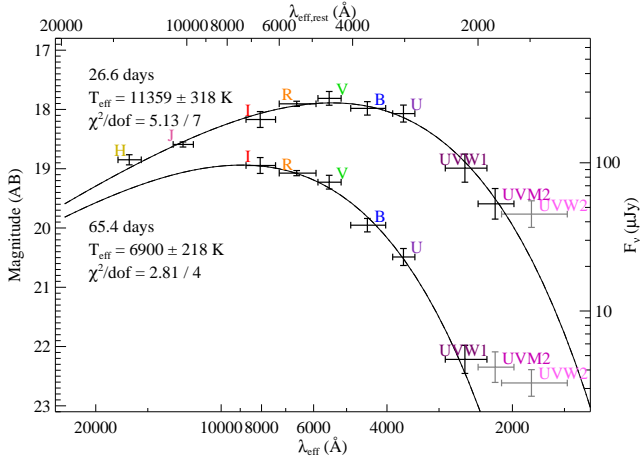,width=3.5in,angle=0}}
\caption{SEDs from \sn\ corrected for Galactic extinction. We show two
representative SEDs of \sn\ at $\sim$27 and $\sim$65 rest-frame days
after discovery. The SEDs are well fit by a single BB component. Note
that relative to the model BB there is excess flux in the $UVW2$ band
in both epochs, while the $UVM2$ band shows excess in the later
epoch. All UV measurements that exhibit a clear excess relative to a
BB are shown in gray and are excluded from the fits.}
\label{SED}
\end{figure}

In Figure~\ref{SED} we show two representative fits to the SED of
\sn\, at 26.6 and 65.4 rest-frame days after discovery. In all epochs
we observe excess emission relative to a single BB in the bluest of
the \swift\ filters, $UVW2$, while starting around 55 rest-frame days
after discovery there is excess emission in both $UVW2$ and
$UVM2$. This excess was also observed in the Type II-Plateau (II-P)
SN~2006bp, where it was attributed to complex line blanketing by
Fe-peak elements \citep{immler07}.  The blue excess is readily
identified in the UV color curves (e.g., $UVW2$--$UVW1$) of \sn, which
evolve toward the red until $\sim$50 rest-frame days after discovery,
at which point the $UVW2$--$UVW1$ and $UVM2$--$UVW1$ colors become
progressively more and more blue.  These points were excluded from our
SED/BB fits. To confirm that we were not underestimating the
bolometric luminosity of the SN, we directly integrated the flux in
the SED using the same method as \citet{modjaz08d}. We show the
results of this direct integration in the lower panel of
Figure~\ref{bol_lum}, and in all cases we find that the BB model and
direct integration method agree to within $\la$20\%.

The observed linear decline in the light curve of \sn\ leads us to
classify it as a Type II-L SN.  We measure a bolometric photometric
decay rate of 0.042 mag d$^{-1}$, which is slightly faster than the
$V$-band (roughly rest-frame $B$) rate of 0.036 mag d$^{-1}$. This
rate is also slightly slower than the rest-frame $B$-band decay of
both SN~1979C (0.046 mag d$^{-1}$; \citealt{panagia1980}) and SN~1980K
(0.055 mag d$^{-1}$; \citealt{barbon82}).  Furthermore, integrating
the bolometric light curve from 15 to 83 rest-frame days after
discovery yields a total radiated energy of $\sim$9 $\times 10^{50}$
ergs, comparable to the canonical 10$^{51}$ ergs deposited into the
kinetic energy of a SN. If we assume the same bolometric correction
factor to the observed $V$-band light curve both pre- and
post-maximum, and we include the early data from \citet{gezari08}, we
find the total radiated energy of \sn\ over the first 83~d after
discovery is $\sim$1.1 $\times 10^{51}$ ergs.

The three \swift\ UVOT observations taken more than 80~d after the
discovery of \sn\ show tentative evidence for a significant reduction
in the SN photometric decline rate (see Figure~\ref{ourphot}). We
note, however, that the late-time UVOT measurements have significant
errors, and these observations may be consistent with the early trend
seen in the light curve. \citet{gezari08} suggested that these
observations were evidence that the SN luminosity made a transition to
being powered by radioactive heating from $^{56}$Co. With only two
observations of this decline, we caution against prematurely
identifying the $^{56}$Co decay tail, and instead argue that the
observations are inconclusive at this time. For instance, it would be
possible to mimic the behavior of $^{56}$Co decay with a late-time
light echo, or interaction with an extended CSM. We note that
SN~2006gy also showed evidence for a ``flattening'' several months
after explosion \citep{smith07-2006gy}, but further observations taken
the following year indicated that this was not clearly the $^{56}$Co
tail \citep{smith08-2006gy}.  Our last optical spectrum
(Fig.~\ref{specevo}) also shows no evidence that the ejecta were
becoming optically thin on that date, as would be expected if \sn\
were transitioning to the radioactive decay tail.  Late-time
photometry will be necessary to conclusively identify if and when the
light curve of \sn\ made a transition to being powered by $^{56}$Co
decay, and thus allow a measurement of the amount of $^{56}$Ni
synthesized in the explosion.

\section{Spectroscopy}

The \sn\ spectral sequence plotted in Figure~\ref{specevo} is labeled
with ages (in the rest frame) relative to the observed $V$-band
maximum in order to facilitate comparison with SNe 1979C and 1980K in
Figure~\ref{speccomp}.  Our first two spectra of \sn\ (at $+3$ and
$+13$ d relative to maximum light) show a smooth and featureless blue
continuum with no identifiable spectral features.  In particular, we
do not detect an emission feature near 5650~\AA\ (4660~\AA\ in the
rest frame) seen in earlier spectra of \sn\ taken between 2008 May 1
and 2008 May 8 \citep{ATEL.1515, gezari08}. \citet{gezari08}
attribute this feature solely to \ion{He}{2} $\lambda$4686, but we
note that transient emission features seen in the early spectra of
some SNe~IIn at similar wavelengths are due to a combination of
\ion{He}{2} $\lambda$4686 and the Wolf-Rayet \ion{C}{3}/\ion{N}{3}
$\lambda$4640 blend \citep{niemela85,leonard00}.

In addition, the spectrum of the extremely luminous SN~2005ap from
$-3$ d is shown at the top of Figure~\ref{specevo}.  The SN~2005ap
spectrum is also very blue, but has additional spectral features that
are not seen in our \sn\ spectrum from a similar
epoch. \citet{quimby05ap} identified the strongest spectral feature in
SN~2005ap, the ``W''-shaped feature near 4200~\AA, as being due to a
blend of \ion{C}{3}, \ion{N}{3}, and \ion{O}{3} with an expansion
velocity of about 20,000 \kms.

Our next spectrum, at $+$21 d, is noticeably redder and is the first
to show strong spectral features.  Both SN~1979C at a similar epoch
and \sn\ show a blue continuum with weak, low-contrast lines of H and
\ion{He}{1} lines present mostly in absorption (Fig.~\ref{speccomp}).
However, H$\alpha$ is present only in emission and is weak in \sn.
The H$\alpha$ line in \sn\ has a full width at half-maximum intensity
(FWHM) of 10,000 \kms\ and an equivalent width of only 22 \AA.  (Both
of these values have large uncertainties due to difficulties in
defining the continuum for a line with such a low amplitude.)  One
difference between the two objects is the lower apparent velocities in
\sn.  The H$\beta$ and \ion{He}{1} $\lambda$5876 lines of SN~1979C
have absorption minima at velocities of 9700 and 8900 \kms.  In \sn,
these values are 6000 and 5700 \kms, respectively, although we caution
that the exact values are correlated with the assumed value for the
redshift.

Over the next two months, the spectra of \sn\ plotted in
Figure~\ref{specevo} became redder, reflecting the cooling
photospheric temperature evolution discussed above. In addition, the
spectral features first seen in the day +21 spectrum gradually became
more prominent.  The SN features are still muted in amplitude relative
to those expected in a normal Type II SN.  This may be an example of
the ``top-lighting'' effect described by \citet{branch00}, where
continuum emission from interaction with CSM illuminates the SN ejecta
from above and results in a rescaling of the amplitudes of spectral
features.

By the time of our day +68 spectrum, the P-Cygni spectral features due
to H Balmer lines, \ion{Na}{1}, and \ion{Fe}{2} become more prominent
and the overall appearance starts to resemble that of normal SNe II.
The H$\alpha$ profile lacks an absorption component, which may be
common to SNe II-L and not SNe II-P (e.g., \citealt{schlegel96},
\citealt{fil97}).  The broad H$\alpha$ emission extends (at
zero intensity) from $-9000$ to 9000 \kms, with a FWHM of 9500 \kms.
This velocity width is intermediate between that of the SNe~1979C and
1980K spectra (FWHM $\approx$ 10,600 and $\approx$ 8200 \kms,
respectively).

\begin{figure}
\centerline{\psfig{file=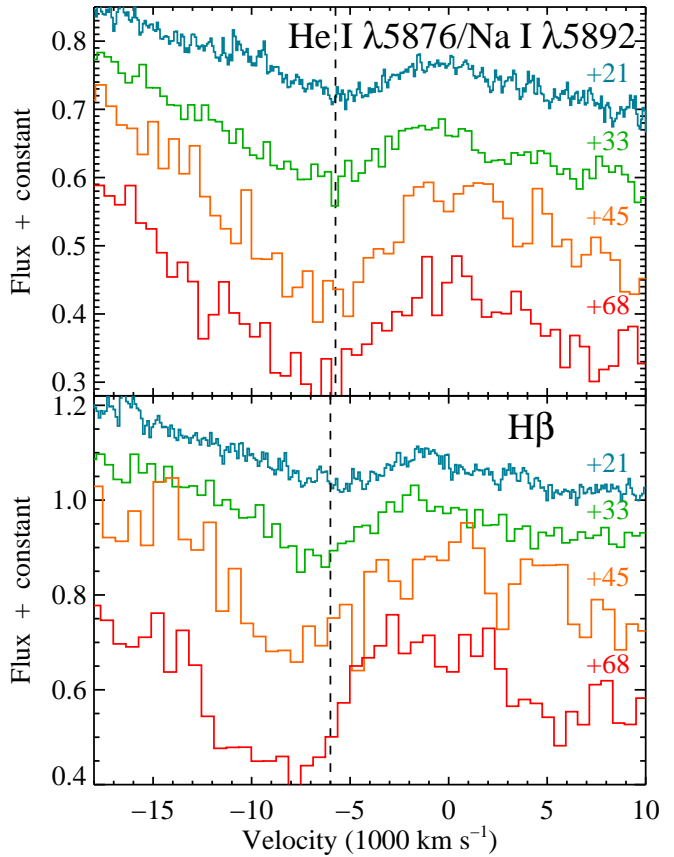,width=3.5in,angle=0}}
\caption{Velocity evolution of absorption minima.  The spectra are
labeled with dates after maximum light in the same manner as in
Figure~\ref{specevo}. In the bottom panel the apparent blueshift of
the H$\beta$ absorption minimum increases over time.  A vertical
dashed line at $-$6000 \kms\ marks the velocity of the absorption
minimum on day 21 to guide the eye.  In the top panel, the evolution
of the \ion{He}{1} $\lambda$5876/\ion{Na}{1} $\lambda$5892 blend is
plotted, with $\lambda$5876 used as the zero point of the velocity
scale.  The vertical dashed line at $-$5700 \kms\ marks the absorption
minimum on day 21 to guide the eye.}
\label{velplot}
\end{figure}

Unlike in most core-collapse SNe, the velocity of the minimum of the
H$\beta$ line {\it increased} over time, from 6000 \kms\ at day 21, to
8700 \kms\ at day 68, as shown in the bottom panel of
Figure~\ref{velplot}.  The exact values of the velocity depend
directly on the assumed redshift, but the {\it trend} is
independent of those uncertainties.  Unfortunately, the other
absorption lines are mostly blended (except for H$\alpha$, which does
not show an absorption component), so we cannot isolate the velocity
trend in other spectral features without spectral modeling.  The top
panel of Figure~\ref{velplot} shows the evolution of the \ion{He}{1}
$\lambda$5876/\ion{Na}{1} $\lambda$5892 blend.  At early times
\ion{He}{1} dominates the blend, and as the ejecta cool at later
epochs \ion{Na}{1} should dominate, resulting in an $\sim$800 \kms\
redward shift of the rest wavelength.  After taking into account that
redward shift, the top panel of Figure~\ref{velplot} also shows some
weak evidence for a blueward shift of the absorption minimum to
$\sim$8000 \kms.

The only supernova known to us to show increasing absorption
velocities over time is the peculiar SN~Ib~2005bf. The trend of
increasing absorption velocities was only visible in the three He~I
lines, but not in other lines such as Ca~II H\&K
\citep{modjaz07_thesis}.  \citet{tominaga05bf} explained the effect as
being due to progressive outward excitation of \ion{He}{1} by
radioactive $^{56}$Ni in the interior as the ejecta expanded and the
density decreased, an effect that seems to be of little relevance to
\sn.  A more likely possibility is that blending with some
unidentified line is shifting the velocity of the apparent absorption
minimum.  At late times H$\beta$ dominates its region of the spectrum,
but at early times \ion{He}{2} $\lambda$4686 could possibly be
contributing emission to the blue wing of the H$\beta$ absorption
profile.  Another scenario is that some unusual, but as yet
unidentified, radiative transfer effect is affecting the wavelength of
the apparent absorption minimum in \sn.  For example, if the optical
depth in an optically thick line increases sufficiently rapidly with
time (e.g., as the ejecta cool and recombine), then the apparent
absorption minimum could move blueward \citep{jb90}.  None of these
suggestions explain why such an effect would be present only in \sn\
and not in normal SNe.

\section{Discussion}

\subsection{The Physical Nature of \sn}

\begin{figure}
\centerline{\psfig{file=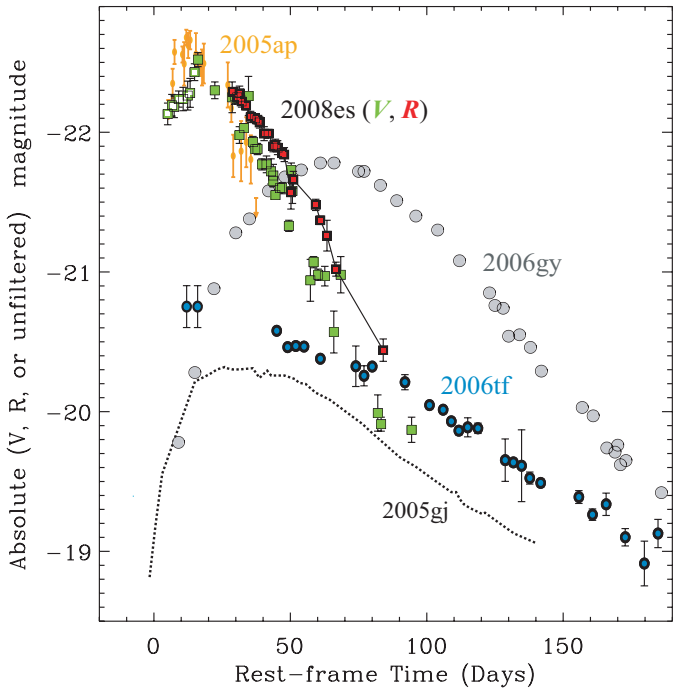,width=3.5in,angle=0}}
\caption{Rest-frame brightness evolution of five of the most luminous
known SNe. For SN~2008es we derive the absolute magnitude using UVOT 
and the KAIT/Nickel $V$-band magnitudes (filled green squares) and the 
KAIT/Nickel $R$-band
magnitudes (red squares), assuming that day 0 is the discovery date. 
Early $g$-band photometry from \citet{gezari08}, converted to 
$V$-band using the color equations of \citet{jester2005}, is also shown 
(open green squares).  The light curve for SN~2005ap (orange) is derived from
unfiltered photometry in \citet{quimby05ap}, and SN~2006gy (gray
circles) is also unfiltered from \citet{smith07-2006gy}.  $R$-band
photometry of SN~2006tf (blue circles) is from \citet{smith06tf}, and
$r$/$r'$ photometry of SN~2005gj (dotted line) is from
\citet{prieto05gj}. The light curve for SN~2006tf is shifted by +16~d 
from that in \citet{smith06tf}; since the explosion date is not
known, we chose to align its time of peak luminosity with those of
SNe~2005ap and 2008es.}
\label{photcomp}
\end{figure}

The photometric evolution of \sn\ (see Fig.~\ref{photcomp}) is much
faster than that of other very luminous SNe like SNe~2006gy
\citep{smith07-2006gy, ofek06gy}, 2006tf \citep{smith06tf}, and 2005gj
\citep{prieto05gj}, and its spectrum also betrays no evidence for the
strong CSM interaction seen in these other SNe~IIn, in the form of
narrow lines from the CSM or intermediate-width H$\alpha$ from the
post-shock shell.  Indeed, considering both its photometric and
spectroscopic evolution, we suggest that \sn\ is most like the
overluminous SN~2005ap \citep{quimby05ap} and thereby most closely
resembles a Type II-L that is 4--5 mag more luminous than typical
SNe~II-L \citep{richardson02}.

To power the tremendous luminosity of \sn\ with radioactive decay
would require an initial $^{56}$Ni mass of $\sim$10 M$_{\odot}$ (see,
e.g., \citealt{smith07-2006gy}).  This $^{56}$Ni mass would very
likely need to be generated in a pair-instability explosion
(\citealt{barkat67,bond84}), but this $^{56}$Ni mass seems problematic
given that it would be larger than the modest envelope mass indicated
by the relatively fast rise and decay time (see below).  In addition,
the photometric decline of 0.042 mag d$^{-1}$ is faster than that of
$^{56}$Co, 0.0098 mag d$^{-1}$, making radioactive heating unlikely
as the dominant source of the energy.

Despite the lack of a Type IIn spectrum, the most likely
interpretation seems to be that the high luminosity of \sn\ is the
result of converting shock energy into visual light.  This can be
accomplished, in principle, if the shock kinetic energy is thermalized
throughout a massive envelope, like a normal SN~II-P or II-L, but with
a much larger initial radius of (2--3) $\times 10^{15}$ cm (based on
the peak luminosity and the evolution in Fig.~\ref{bol_lum}). Were
the initial radius much smaller than this it would prove difficult to
convert $\sim$10$^{51}$ ergs of kinetic energy to $\sim$10$^{51}$ ergs
of radiation because there would be significant adiabatic losses.  The
apparent temperature evolution from $\sim$15,000 to 6000 K over a time
period of $\sim$66~d (Fig.~\ref{bol_lum}), reminiscent of other
normal SNe~II, suggests that the recombination photosphere is receding
through a cooling envelope in \sn.  In this scenario, the usual
narrow/intermediate-width H$\alpha$ emission that is taken as a
signpost of CSM interaction might be avoided if the CSM shell is
initially very opaque, and if the shock encounters no further CSM
at larger radii (see \citealt{smith06tf}).

Since the required initial radius exceeds that of the largest known
red supergiants (see \citealt{smith2001}) by a factor of 20--30, it
requires the envelope to be an unbound, opaque CSM shell ejected prior
to the SN explosion instead of a traditional bound stellar envelope.
Similar models were suggested for SN~2005ap \citep{quimby05ap},
SN~2006gy \citep{smith+mccray07}, and SN~2006tf \citep{smith06tf},
implying CSM envelope masses of 0.6, $\sim$10, and 18 M$_{\odot}$,
respectively. The corresponding CSM mass for \sn\ would be roughly
$\sim$5 M$_{\odot}$ in this scenario, because its evolution and
expansion speeds are slower than those of SN~2005ap.  This CSM mass of
$\sim$5~M$_{\odot}$ allows the observed H$\alpha$ Doppler velocities
to remain faster than in SNe~2006gy and 2006tf, where the heavy CSM
shells decelerated the shocks to only 4000 and 2000 km s$^{-1}$,
respectively (\citealt{smith07-2006gy}; \citealt{smith06tf}). The
smaller CSM mass for \sn\ relative to SNe~2006gy and 2006tf would also
explain the comparatively short rise and decay observed for \sn, as
the light diffusion time for this SN is much shorter than in
SNe~2006gy and 2006tf \citep{smith+mccray07}.  The putative envelope
ejection preceding \sn\ must have occurred 10--100 yr prior to the
explosion (for an unknown progenitor wind speed of $v_{\rm CSM}$ =
10--100 km s$^{-1}$), indicating a progenitor mass-loss rate of order
0.05--0.5 M$_{\odot}$ yr$^{-1}$.  This is much larger than any steady
stellar wind (see \citealt{smith07-2006gy}, and references therein),
providing another case of impulsive mass ejection in the decades
immediately preceding some SNe.

An alternative analysis of \sn, submitted for publication shortly
after this paper was made available electronically, also suggests 
that the extreme luminosity is powered via interaction with CSM
\citep{gezari08}. However, the initial version of \citet{gezari08}
argues against an unbound CSM shell ejection prior to the SN
explosion, and instead argues that the CSM is created by a dense 
progenitor wind.

\subsection{The Host of \sn}

Currently there is no conclusive detection of the host galaxy of
SN~2008es, so its metallicity and corresponding implications for the
pre-SN evolution are not known. From SDSS
DR6 images \citep{adelman08}, we derive a 3$\sigma$ upper limit of
$m_{r'} >$ 22.7 mag at the SN position, which translates to roughly
$M_V > -$17.4 mag at the SN redshift (neglecting any K
corrections). Thus, the putative host galaxy is significantly less
luminous than an $L_*$ galaxy and could be comparable to or fainter
than the Small Magellanic Cloud (with $M_V=-16.9$ mag)\footnote{A
galaxy $\sim$9$\arcsec$ to the NE of the SN position is unlikely to be
the host itself even if found to be at a similar redshift as \sn; this
would require significant massive star formation at a projected
physical distance of $\sim$31 kpc.}. 

Without a definitive detection of the host galaxy we assume that there
is no host extinction. This assumption is further supported by both
the low luminosity of the host galaxy and our BB fits (see $\S$3) that
show excellent agreement with our UV measurements, where the effects
of host extinction would be especially prominent. In analogy with the
underluminous hosts of SNe~2006tf \citep{smith06tf} and 2005ap
\citep{quimby05ap}, deep imaging after the SN has faded may uncover
the host galaxy nearly coincident with the SN position, which would
allow further constraints to be placed on extinction in the
host. These three extreme SNe and their underluminous hosts may be
hinting that very luminous SNe II preferentially occur in
low-luminosity host galaxies.

\subsection{Rates of Extremely Luminous \sn-like Events}

Over the past three years the TSS has successfully found tens of SNe
with a surprising rate of unusual objects. The list of such SNe
includes SN~2005ap, SN~2006gy, SN~2006tf, SN~2008am, and now
\sn. However, this apparent high anomaly rate is probably the result
of combining a huge survey volume with an intrinsically rare class of
objects. Following \citet{quimby2008}, we compare the volume probed by
the TSS for SNe Ia, and bright core-collapse SNe. The highest redshift
of the $\sim$30 SNe~Ia found by the TSS is $z\approx0.1$, while
SN~2005ap was found at $z\approx0.3$. The comoving volume scanned by
the TSS is therefore $\sim$23 times bigger for SN~2008es-like objects
than for SNe~Ia. Of the five bright objects found by the TSS, only
two, SN~2005ap and SN~2008es, can be considered together as a
class. Thus, the comparative rate is about $30/2 \times 23 \approx
350$ times smaller.

The KAIT SN search \citep{filippenkoLOSS} has discovered about 400
SNe~Ia in the past 10 years in targeted nearby galaxies, hence the
expected number of SN~2008es-like objects is of order 1. This is
consistent with the nondetection of any such SN.

Using the light curve of \sn\ and the accumulated observations of
about $10,000$ spiral galaxies in the KAIT sample which were observed
regularly during the past ten years, we find an upper limit on the
rate (per unit $K$-band luminosity) of such SNe to be about $1/160$
times the local SN~II rate \citep{leaman2009}. Furthermore, we note
that if SN~2008es-like events preferentially occur in small,
low-metallicity galaxies (see $\S$ 5.2), the KAIT sample, which
targets large, nearby galaxies, would be biased against the discovery
of such a SN.  While there are several SN search surveys with larger
search volumes than the TSS, such as the SDSS-II Supernova Survey
\citep{sako08}, the full details of these surveys (cadence, total
survey volume, candidate SN color cuts, etc.) are not available at
this time, and therefore we cannot directly compare them to the TSS.
We conclude that the number of luminous Type II-L SNe, as discovered
by the TSS, does not appear to be the result of a statistical fluke,
and predict many such detections with future wide-field synoptic
surveys.

\section{Conclusions}

We have reported on our early observations of \sn, which at a peak
optical magnitude of $M_V = -22.3$ is the second most luminous SN ever
observed.  We argue that the extreme luminosity of this SN was likely
powered via strong interaction with a dense CSM, and that the steep
decline in the light curve, 0.042 mag d$^{-1}$, indicates that the
radioactive decay of $^{56}$Co is likely not the dominant source of
energy for this SN. Integration of the bolometric light curve of \sn\
yields a total radiated energy output of $\ga$ 10$^{51}$ ergs.  The
optical spectra of \sn\ resemble those of the luminous SN~1979C, but
with an unexplained increase in the velocity of the H$\beta$
absorption minimum over time. 

 We also examined the rate of discovery of extremely luminous SNe by
the Texas Supernova Search and find that their discovery of the five
most luminous observed SNe in the past four years is probably not a
fluke; several more such detections are expected in the coming years.

Finally, what behavior can we expect from \sn\ at late times, roughly
1 yr or more after discovery?  Regardless of whether the peak
luminosity was powered by radioactive decay or optically thick CSM
interaction (see \citealt{smith08-2006gy}), a SN can be powered by
strong CSM interaction at late times if the progenitor had a
sufficiently high mass-loss rate in the centuries before exploding.
We have seen examples of both: SN~2006tf had strong H$\alpha$ emission
indicative of ongoing CSM interaction at late times \citep{smith06tf},
whereas SN~2006gy did not \citep{smith08-2006gy}.  SN~1979C was less
luminous at peak than those two SNe, but it has been studied for three
decades because its late-time CSM interaction is powering ongoing
emission in the radio, optical, and X-rays \citep{weiler81,fesen93,
immler98}. With such an extraordinarily high peak luminosity, a
late-time IR echo such as that seen in SN~2006gy \citep{smith08-2006gy}
is also likely if \sn\ has dust waiting at a radius of $\sim$0.3 pc.
Alternatively, if \sn\ were powered in whole or in part by
radioactivity, a large mass of $^{56}$Ni should be evident in the
late-time decline rate.

\acknowledgments

We thank the anonymous referee for comments that helped improve this
paper. We are grateful to B.~Jannuzi for approving Kitt Peak DD
observations, and to Diane Harmer and David L. Summers for carrying
out the observations. M. Malkan accommodated a small telescope time
trade, while N. Joubert and B. Macomber assisted with some of the
observations. P. Nugent kindly checked for pre-discovery images of SN
2008es from DeepSky. A.A.M. is supported by a UC Berkeley Chancellor's
Fellowship. M.M. is supported by a Miller Institute research
fellowship. N.R.B. and D.A.P. are partially supported by a SciDAC
grant from the Department of Energy. J.S.B.'s group is partially
supported by NASA/{\it Swift} grant \#NNG05GF55G and a Hellman Faculty
Award.  A.V.F.'s group is supported by National Science Foundation
(NSF) grant AST--0607485 and the TABASGO Foundation.  The Peters
Automated Infrared Imaging Telescope is operated by the Smithsonian
Astrophysical Observatory (SAO) and was made possible by a grant from
the Harvard University Milton Fund, the camera loan from the
University of Virginia, and the continued support of the SAO and UC
Berkeley. The PAIRITEL project is partially supported by NASA/{\it
Swift} Guest Investigator Grant \#NNG06GH50G.  KAIT and its ongoing
operation were made possible by donations from Sun Microsystems, Inc.,
the Hewlett-Packard Company, AutoScope Corporation, Lick Observatory,
the NSF, the University of California, the Sylvia \& Jim Katzman
Foundation, and the TABASGO Foundation. We acknowledge the use of
public data from the {\it Swift} data archive. Some of the data
presented herein were obtained at the W. M. Keck Observatory, which is
operated as a scientific partnership among the California Institute of
Technology, the University of California, and NASA; the Observatory
was made possible by the generous financial support of the W. M. Keck
Foundation. The authors wish to recognize and acknowledge the very
significant cultural role and reverence that the summit of Mauna Kea
has always had within the indigenous Hawaiian community; we are most
fortunate to have the opportunity to conduct observations from this
mountain.



\clearpage

\end{document}